\documentclass{article}
\usepackage{graphicx}

\newenvironment{pf}{\unskip{\bf Proof:}}{\unskip{\hfill $\Box$}}

\newcommand{\lemlab}[1]{\label{lemma:#1}}

\newcommand{\figlab}[1]{\label{fig:#1}}

\newcommand{\figref}[1]{\ref{fig:#1}}

\newtheorem{theorem}{Theorem}[section]
\newtheorem{lemma}[theorem]{Lemma}

{\catcode`\@=11
\gdef\setft#1#2#3{%
\def\@oddfoot{
{\setbox0=\hbox{#1}
\setbox1=\hbox{#3}
\ifdim\wd0>\wd1
\dimen0=\wd0
\box0\hfil#2\hfil\hbox to\dimen0{\hfil\hfil\box1}
\else \dimen0=\wd1
\hbox to\dimen0{\box0\hfil }\hfil#2\hfil\box1 \fi
}}} }

\def\complaint#1{}
\def\withcomplaints{
\newcounter{mycomplaints}
\def\complaint##1{\refstepcounter{mycomplaints}%
\ifhmode%
\unskip%
{\dimen1=\baselineskip \divide\dimen1 by 2 %
\raise\dimen1\llap{\tiny -\themycomplaints-}}\fi%
\marginpar{\tiny [\themycomplaints]: ##1}}%
}

\usepackage{amssymb}
\usepackage{amsmath}
\usepackage{latexsym}
\newcommand\R{\mathbb{R}}

\title{Computational Geometry Column 41}
\author{%
Joseph O'Rourke\thanks{
Dept. of Computer Science, Smith Col\-lege, North\-ampton, 
MA 01063, USA.
\-orourke@cs.\-smith\-.edu.
Supported by NSF Grant CCR-9731804.
}
}
\date{}
\begin{document}
\maketitle
\pagestyle{empty}
\thispagestyle{empty}

\begin{abstract}
The recent result that $n$ congruent balls in $\R^d$ have
at most $4$ distinct geometric permutations is described.
\end{abstract}

A line $\ell$ {\em stabs\/} a set $S$ of geometric objects in $\R^d$
if $\ell$ intersects every member of $S$.
Such a stabber is often called 
a {\em line traversal\/} of $S$.
When the objects are pairwise disjoint convex bodies, then
a line traversal induces a {\em geometric permutation\/} of the objects:
either of the two piercing orders of the objects 
(one the reverse of the other).
Basic bounds on the number of geometric permutations were established
about a decade ago: the number is in
$\Omega(n^{d-1})$~\cite{kll-dwsdc-92}
and
$O(n^{2d-2})$~\cite{w-ubgpc-90}, 
with a tight bound of $2n-2$ known for $d=2$~\cite{es-mnwsn-90}.
The study of geometric permutations was 
revitalized by a focus
specifically on balls by
Smorodinsky, Mitchell, and Sharir~\cite{sms-sbgp-00}.
They closed the gap in this special case to $\Theta(n^{d-1})$,
a result which was quickly followed by an extension to the same
bound for ``fat'' convex objects~\cite{kv-tbngp-99}.

\begin{figure}[htbp]
\centering
\includegraphics[width=0.9\linewidth]{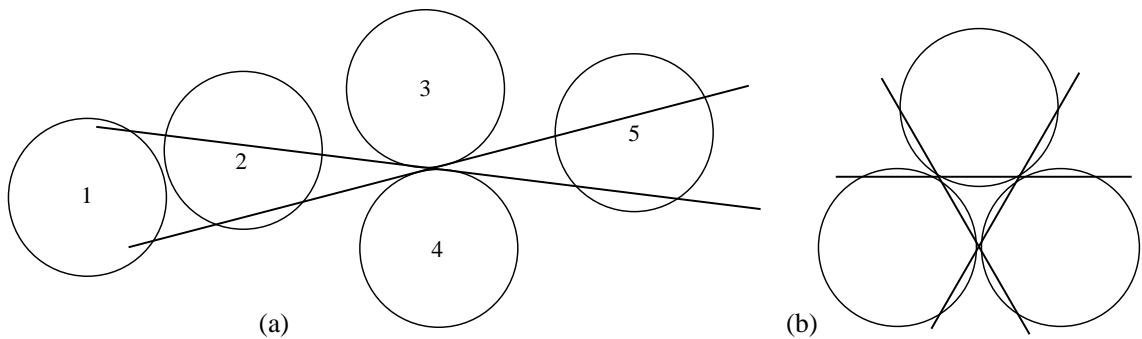}
\caption{(a) Two line traversals with permutations $(1,2,3,4,5)$ and $(1,2,4,3,5)$;
(b) Three permutations achievable.}
\figlab{disks.stabbed}
\end{figure}
One fascinating line of investigation opened in~\cite{sms-sbgp-00},
and independently by Asinowski and Katchalski~\cite{a-ctgp-98},
is the even more special case of congruent balls.
They proved that sufficiently many congruent disks in $\R^2$ 
admit only a constant number of geometric permutations---in fact,
just $2$.
A little experimentation (Fig.~\figref{disks.stabbed}a) quickly
reveals the operable constraint:  one line traversal necessarily strings
out the disks roughly along a line, and any other stabbing line
cannot deviate too much if it is to pierce all the disks.
The only example that admits more than $2$
permutations is a triangular arrangment of three disks 
(Fig.~\figref{disks.stabbed}b); 
the proof in~\cite{a-ctgp-98} shows that the bound of $2$ holds
for all $n \ge 4$.
If the disks are not congruent, then $\Omega(n)$ geometric permutations
are possible.

Smorodinsky et al. left open the question of bounds for congruent
balls in $d \ge 3$ dimensions, but conjectured that again only $O(1)$
geometric permutations are achievable.
This conjecture was settled by Zhou and Suri,
who established an upper bound of $16$ for all $d$, for a sufficiently
large number $n$ of balls.
Quickly they~\cite{zs-ssgp-01}, 
and independently Huang, Xu, and Chen~\cite{hxc-gphds-01},
improved this bound to $4$, where it stands at this writing.
Both sets of researchers have also extended their results to
noncongruent balls of bounded radius ratio, delimiting when
the $n^{d-1}$ behavior kicks in.

A sense for the congruent-balls problem can be obtained by attempting to stab
four spheres stacked as cannonballs
(displaced slightly to ensure pairwise disjointness).
See Fig.~\figref{balls}a.
There is no line traversal for this set of balls,
although I do not know a simple proof of this claim.
Again, the existence of
a line traversal $\ell$ requires the balls to be strung out along $\ell$,
a requirement incompatible with the clumping in this stack.
\begin{figure}[htbp]
\centering
\includegraphics[width=0.6\linewidth]{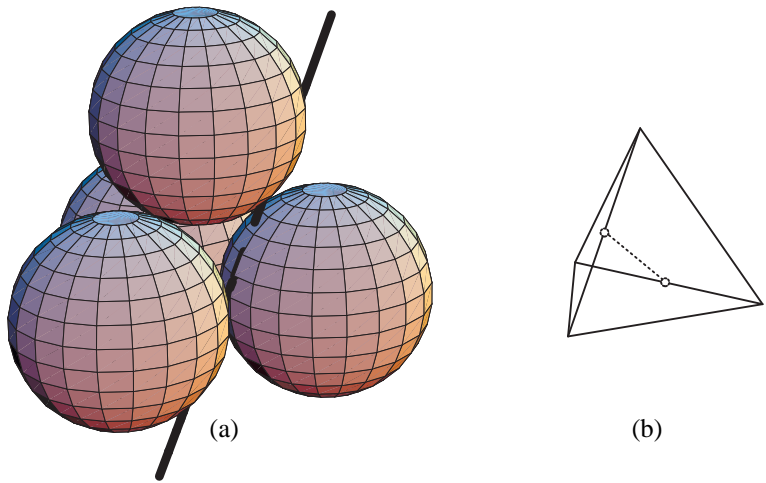}
\caption{(a) A stabbing of three out of four congruent balls;
(b) The centroids of the two pairs of unit-radius
balls are separated by 
$\sqrt {2}$ 
(dashed segment).
}
\figlab{balls}
\end{figure}

I now sketch the proof of
Zhou and Suri for a set $S$ of unit-radius balls
in $\R^d$, relying on~\cite{z-ssgc-00}.
They call a pair of balls in the set $S$
a {\em switched pair\/}
if, like disks $3$ and $4$ in
Fig.~\figref{disks.stabbed}a, 
there are line traversals for $S$ 
that meet them in two different orders.
They first prove that the balls in any switched pair must be close
to one another, and next, that a stabber $\ell$ of such
a switched pair must be nearly perpendicular to the line
through the centers of the two balls,
and pass very nearly through their common
center of gravity (centroid).
(These constraints are evident in Fig.~\figref{disks.stabbed}a.)

From this they prove
that each ball of $S$ can participate
in at most one switched pair, which in turn
leads to the conclusion that the two balls in a switched pair
must appear consecutively in every geometric permutation of $S$.

The theorem is established by computing upper and
lower bounds on the ``separation'' between two distinct switched pairs.
They prove the distance between the centers of gravity
of two switched pairs has a lower bound
of $\sqrt{2} - \epsilon(n)$, where $\epsilon(n)>0$
and $\lim_{n \rightarrow \infty} \epsilon(n) = 0$.
This $\sqrt{2}$ term can be seen to derive
(roughly) from the distance between the centers of gravity of the
two pairs of balls in the cannonball stack of Fig.~\figref{balls}b:
this is the distance between the midpoints of opposite edges of
the tetrahedron formed by the ball centers.

So switched pairs cannot be too close together.
Neither can they be too far apart, a claim we will leave to
the intuition that the need for stabbers to pass close to the
centers of gravity of the switched pairs limits their separation.
Combining the explicit bounds leads to the conclusion
that there can be at most two switched pairs, which in
turn establishes a bound of $4$ on the number of geometric
permutations for sufficiently large $n$.

It remains possible that the correct bound is $2$,
again for sufficiently large $n$.
For small values of $n$,
examples might exist that determine more than $4$ permutations.
But as far as I know,
no one has found a collection of $n \ge 4$ congruent
balls in $\R^3$ that have more than one switched
pair, or admit more than $2$ geometric permutations.

Although the pursuit of geometric permutations has been largely
driven by theoretical interests, there is an important application in
$\R^3$ to computer graphics: lines of sight between mutually
visible objects represent stabbers.  Bounds on the number
of stabbers can translate into bounds on visibility complexity.
This makes especially interesting the
result in~\cite{zs-ssgp-01} that a collection of pairwise
disjoint, axis-aligned boxes (e.g., the ``bounding boxes'' so useful
in graphics)
admit only $2^{d-1}$ geometric permutations, i.e., $4$ in $\R^3$.
This bound is tight.

\bibliographystyle{alpha}
\bibliography{41}
\end{document}